\documentclass[preprint]{ptephy}

\preprintnumber{KYUSHU-HET-154}

\usepackage{amssymb}
\usepackage{amsthm}
\usepackage{amsmath}
\usepackage{booktabs}
\DeclareMathOperator{\tr}{tr}

\newcommand{\Slash}[1]{{\ooalign{\hfil/\hfil\crcr$#1$}}}
\numberwithin{equation}{section}

\usepackage{url}

\begin{document}

\title{Background field method in the gradient flow}

\author{\name{\fname{Hiroshi} \surname{Suzuki}}{1,\ast}
}

\address{%
\affil{1}{Department of Physics, Kyushu University, 6-10-1 Hakozaki, Higashi-ku, Fukuoka, 812-8581, Japan}
\email{hsuzuki@phys.kyushu-u.ac.jp}
}

\begin{abstract}
In perturbative consideration of the Yang--Mills gradient flow, it is useful
to introduce a gauge non-covariant term (``gauge-fixing term'')  to the flow
equation that gives rise to a Gaussian damping factor also for gauge degrees of
freedom. In the present paper, we consider a modified form of the gauge-fixing
term that manifestly preserves covariance under the background gauge
transformation. It is shown that our gauge-fixing term does not affect
gauge-invariant quantities as the conventional gauge-fixing term. The
formulation thus allows a background gauge covariant perturbative expansion of
the flow equation that provides, in particular, a very efficient computational
method of expansion coefficients in the small flow time expansion. The
formulation can be generalized to systems containing fermions.
\end{abstract}
\subjectindex{B01, B31, B32, B38}
\maketitle

\section{Introduction}
\label{sec:1}
As a novel method to define renormalized quantities, the Yang--Mills gradient
flow~\cite{Luscher:2010iy,Luscher:2011bx} and its extension to the fermion
field~\cite{Luscher:2013cpa} have attracted much attention in recent years,
mainly in the context of lattice gauge theory. Reference~\cite{Ramos:2015dla}
is a recent review,
and~Refs.~\cite{McGlynn:2014bxa,Bruno:2014ova,Christensen:2014raa,%
Kikuchi:2014rla,Ramos:2014kla,Chowdhury:2014mra,Brower:2014dfa,Bruno:2014ufa,%
Bruno:2014jqa,Bergner:2014ska,Blum:2014tka,Jin:2014hea,Perez:2014isa,%
Athenodorou:2014eua,Aoki:2014dxa,Monahan:2015lha,Guo:2015tla,Fodor:2015baa,%
Asakawa:2015vta,Perez:2015ssa,Ce:2015qha,Fodor:2015zna} are more recent related
studies.

Although the gradient flow in lattice gauge theory is utilized to study
non-perturbative dynamics of gauge theory, information available through
perturbative theory is always useful because the latter is well under analytic
control. In the present paper, aiming at possible simplification in
perturbative calculations associated with the gradient flow, we consider the
application of the idea of the background field method~\cite{'tHooft:1975vy,%
DeWitt:1980jv,Boulware:1980av,Abbott:1980hw,Ichinose:1981uw} to the gradient
flow. It is well known that this method considerably simplifies perturbative
computation of, e.g., renormalization constants.

As clarified in~Ref.~\cite{Luscher:2010iy}, for perturbative consideration of
the gradient flow, it is useful to introduce a ``gauge-fixing term'' that
breaks gauge covariance of the flow equation; this term gives rise to a
Gaussian damping factor also for gauge degrees of freedom and ensures a
convergence property of momentum integrals. Here, we consider a modified form
of the gauge-fixing term in the flow equation that manifestly preserves
covariance under the background gauge transformation. It is shown that our
gauge-fixing term does not affect gauge-invariant quantities, as the
conventional gauge-fixing term. This formulation thus allows a background gauge
covariant perturbative expansion of the flow equation that provides, in
particular, a very efficient computational method of expansion coefficients in
the small flow time expansion~\cite{Luscher:2011bx}.

This paper is organized as follows. In~Sect.~\ref{sec:2}, we present our
general formulation. Both flow equations for the gauge field and for the
fermion fields are considered. The most important observation is the
independence of gauge-invariant quantities on the gauge-fixing term we
introduce (Sect.~\ref{sec:2.3}). In subsequent sections, we consider
applications of the formulation: In~Sect.~\ref{sec:3}, we consider the
computation of expansion coefficients in the small flow time
expansion~\cite{Luscher:2011bx} relevant to the construction of the lattice
energy--momentum tensor; this computation was carried out
in~Ref.~\cite{Suzuki:2013gza} using a cumbersome diagrammatic method. We
observe that the application of our formulation provides a very efficient
non-diagrammatic computational method, that is quite analogous to that
of~Ref.~\cite{Fujikawa:1993xv}, for the expansion
coefficients.\footnote{Unfortunately, the results of this new simple
computational scheme do not coincide with the results
in~Ref.~\cite{Suzuki:2013gza}, revealing that there are errors in the one-loop
diagrammatic calculation in~Ref.~\cite{Suzuki:2013gza}. The diagrams in which
the mistakes were made in~Ref.~\cite{Suzuki:2013gza} have been completely
identified. For corrected results, see the errata
for~Refs.~\cite{Suzuki:2013gza,Makino:2014taa}
and~Refs.~\cite{Suzuki:2013gzaa,Makino:2014taaa}.} In~Sect.~\ref{sec:4}, we
consider the small flow time expansion relevant to the construction of the
axial-vector current~\cite{Endo:2015iea}. The last section is devoted to the
conclusion.

Here is a summary of our notation: Our generators~$T^a$ of the gauge group~$G$
are anti-Hermitian and the structure constants are defined
by~$[T^a,T^b]=f^{abc}T^c$. Quadratic Casimirs are defined
by~$f^{acd}f^{bcd}=C_2(G)\delta^{ab}$ and, for a representation~$R$,
$\tr_R(T^aT^b)=-T(R)\delta^{ab}$ and~$T^aT^a=-C_2(R)1$. We also denote
$\tr_R(1)=\dim(R)$. For example, for the fundamental $N$~representation
of~$SU(N)$ for which $\dim(N)=N$, the conventional choice is
\begin{equation}
   C_2(SU(N))=N,\qquad T(N)=\frac{1}{2},\qquad
   C_2(N)=\frac{N^2-1}{2N}.
\label{eq:(1.1)}
\end{equation}

Our gamma matrices are Hermitian and for the trace over the spinor index we set
$\tr(1)=4$ for any spacetime dimension~$D$. The chiral matrix is defined
by~$\gamma_5=\gamma_0\gamma_1\gamma_2\gamma_3$ for any~$D$ and thus
\begin{equation}
   \tr(\gamma_5\gamma_\mu\gamma_\nu\gamma_\rho\gamma_\sigma)
   =\begin{cases}
   4\epsilon_{\mu\nu\rho\sigma},&\mu,\nu,\rho,\sigma\in\{0,1,2,3\},\\
   0,&\text{otherwise},
   \end{cases}
\label{eq:(1.2)}
\end{equation}
where the totally anti-symmetric tensor is normalized as~$\epsilon_{0123}=1$.

\section{Flow equations with a background covariant gauge}
\label{sec:2}
\subsection{Gradient flow equation with a background covariant gauge}
The gradient flow for the gauge potential is defined by~\cite{Luscher:2010iy}
\begin{equation}
   \partial_tB_\mu(t,x)=D_\nu G_{\nu\mu}(t,x)
   +\alpha_0D_\mu\partial_\nu B_\nu(t,x),\qquad
   B_\mu(t=0,x)=A_\mu(x),
\label{eq:(2.1)}
\end{equation}
where $t\in[0,\infty)$, and
\begin{equation}
   D_\mu=\partial_\mu+[B_\mu,\cdot],\qquad
   G_{\mu\nu}(t,x)
   =\partial_\mu B_\nu(t,x)-\partial_\nu B_\mu(t,x)
   +[B_\mu(t,x),B_\nu(t,x)]
\label{eq:(2.2)}
\end{equation}
denote the covariant derivative and the field strength of the flowed gauge
field, respectively. The last term in the first relation
of~Eq.~\eqref{eq:(2.1)} breaks gauge covariance and here it is referred to as a
``gauge-fixing term''. As noted in~Ref~\cite{Luscher:2010iy}, for perturbative
consideration of the gradient flow, such as that in~Ref.~\cite{Luscher:2011bx},
it is useful to introduce such a gauge-breaking term because it gives rise to a
Gaussian damping factor also for gauge degrees of freedom and ensures a
convergence property of momentum integrals. It can, however, be shown
that~\cite{Luscher:2010iy} any gauge-invariant quantity, that does not contain
the flow time derivative~$\partial_t$, is independent of the
``gauge parameter''~$\alpha_0$ and physical observables are not affected by the
gauge-fixing term.

In the present paper, we propose a slight modification of the gauge-fixing term
in~Eq.~\eqref{eq:(2.1)}. First, following the general idea of the background
field method~\cite{'tHooft:1975vy,DeWitt:1980jv,Boulware:1980av,Abbott:1980hw,%
Ichinose:1981uw}, we decompose the original gauge potential into the background
part~$\Hat{A}_\mu(x)$ and the quantum part~$a_\mu(x)$ as
\begin{equation}
   A_\mu(x)=\Hat{A}_\mu(x)+a_\mu(x).
\label{eq:(2.3)}
\end{equation}
We also decompose the flowed gauge potential~$B(t,x)$ into the background
part~$\Hat{B}_\mu(t,x)$ and the quantum part~$b_\mu(t,x)$ as
\begin{equation}
   B_\mu(t,x)=\Hat{B}_\mu(t,x)+b_\mu(t,x).
\label{eq:(2.4)}
\end{equation}
Then, our proposal is to adopt, instead of~Eq.~\eqref{eq:(2.1)},
\begin{equation}
   \partial_tB_\mu(t,x)=D_\nu G_{\nu\mu}(t,x)
   +\alpha_0D_\mu\Hat{D}_\nu b_\nu(t,x),\qquad
   B_\mu(t=0,x)=A_\mu(x),
\label{eq:(2.5)}
\end{equation}
where
\begin{equation}
   \Hat{D}_\mu=\partial_\mu+[\Hat{B}_\mu,\cdot]
\label{eq:(2.6)}
\end{equation}
are the covariant derivatives with respect to the \emph{background flowed
field}.

As a further natural assumption, we suppose that the background field is
evolved by its own flow equation:
\begin{equation}
   \partial_t\Hat{B}_\mu(t,x)=\Hat{D}_\nu\Hat{G}_{\nu\mu}(t,x),\qquad
   \Hat{B}_\mu(t=0,x)=\Hat{A}_\mu(x),
\label{eq:(2.7)}
\end{equation}
where
\begin{equation}
   \Hat{G}_{\mu\nu}(t,x)
   =\partial_\mu\Hat{B}_\nu(t,x)-\partial_\nu\Hat{B}_\mu(t,x)
   +[\Hat{B}_\mu(t,x),\Hat{B}_\nu(t,x)]
\label{eq:(2.8)}
\end{equation}
is the field strength of the background field.

\subsection{Covariance under the background gauge transformation}
The original gauge transformation
\begin{align}
   A_\mu(x)\to A_\mu(x)+D_\mu\omega(x),
\label{eq:(2.9)}
\end{align}
may be decomposed into the background part and the quantum part; how this
decomposition is made is the heart of the background field
method~\cite{'tHooft:1975vy,DeWitt:1980jv,Boulware:1980av,Abbott:1980hw,%
Ichinose:1981uw}. A fundamental notion is the \emph{background gauge
transformation}, defined by
\begin{align}
   \Hat{A}_\mu(x)\to\Hat{A}_\mu(x)+\Hat{D}_\mu\omega(x),\qquad
   a_\mu(x)\to a_\mu(x)+[a_\mu(x),\omega(x)].
\label{eq:(2.10)}
\end{align}
The sum of these two reproduces the original gauge
transformation~\eqref{eq:(2.9)}. Under this background gauge transformation,
the quantum gauge field transforms as the adjoint representation. This
transformation can also be generalized to the flowed fields as\footnote{The
covariant derivative~$\Hat{D}_\mu$ in the first relation is defined with
respect to the flowed background field~$\Hat{B}_\mu(t,x)$.}
\begin{equation}
   \Hat{B}_\mu(t,x)\to\Hat{B}_\mu(t,x)+\Hat{D}_\mu\omega(x),\qquad
   b_\mu(t,x)\to b_\mu(t,x)+[b_\mu(t,x),\omega(x)].
\label{eq:(2.11)}
\end{equation}
Note that here we are assuming that the transformation function~$\omega(x)$
does not depend on the flow time~$t$.

Since $\Hat{D}_\mu$ in~Eq.~\eqref{eq:(2.6)} transforms covariantly under the
background gauge transformation~\eqref{eq:(2.11)}, our flow
equation~\eqref{eq:(2.5)} transforms covariantly under the background gauge
transformation; fields transformed by the background gauge transformation obey
the identical equation. This is the reason for our choice of the particular
gauge-fixing term in~Eq.~\eqref{eq:(2.5)} instead of the conventional one
in~Eq.~\eqref{eq:(2.1)}.

Now let us confirm that our gauge-fixing term in~Eq.~\eqref{eq:(2.5)} does not
affect gauge-invariant quantities.

\subsection{Independence of gauge-invariant quantities of the gauge parameter~$\alpha_0$}
\label{sec:2.3}
Although our gauge-fixing term $\alpha_0D_\mu\Hat{D}_\nu b_\nu(t,x)$
in~Eq.~\eqref{eq:(2.5)} differs from the conventional one
in~Eq.~\eqref{eq:(2.1)}, one can still see that any gauge-invariant
quantity, that does not contain the flow time derivative~$\partial_t$ is
independent of the ``gauge parameter''~$\alpha_0$; the gauge-fixing term thus
does not affect gauge-invariant quantities.

To see this, we consider the following ``quantum gauge
transformation''\footnote{Here the covariant derivative~$D_\mu$ in the second
expression is defined with respect to the flowed field~$B_\mu(t,x)$.}
\begin{equation}
   \Hat{B}_\mu(t,x)\to\Hat{B}_\mu(t,x),\qquad
   b_\mu(t,x)\to b_\mu(t,x)+D_\mu\omega(t,x),
\label{eq:(2.12)}
\end{equation}
whose transformation function~$\omega(t,x)$ \emph{does\/} depend on the flow
time~$t$. Note that the sum of these two reproduces the original gauge
transformation~\eqref{eq:(2.9)} with~$\omega(x)\to\omega(t,x)$. Under this
infinitesimal transformation, we have
\begin{align}
   \partial_tB_\mu(t,x)
   &\to\partial_tB_\mu(t,x)+[\partial_tB_\mu(t,x),\omega(t,x)]
   +D_\mu\partial_t\omega(t,x),
\label{eq:(2.13)}
\\
   D_\nu G_{\nu\mu}(t,x)
   &\to D_\nu G_{\nu\mu}(t,x)+[D_\nu G_{\nu\mu}(t,x),\omega(t,x)],
\label{eq:(2.14)}
\\
   \Hat{D}_\nu b_\nu(t,x)
   &\to\Hat{D}_\nu b_\nu(t,x)+\Hat{D}_\nu D_\nu\omega(t,x),
\label{eq:(2.15)}
\end{align}
and
\begin{equation}
   D_\mu\Hat{D}_\nu b_\nu(t,x)
   \to D_\mu\Hat{D}_\nu b_\nu(t,x)
   +[D_\mu\Hat{D}_\nu b_\nu(t,x),\omega(t,x)]
   +D_\mu D_\nu\Hat{D}_\nu\omega(t,x),
\label{eq:(2.16)}
\end{equation}
where in deriving the last expression we have noted
\begin{equation}
   \Hat{D}_\nu D_\nu\omega(t,x)
   =D_\nu\Hat{D}_\nu\omega(t,x)+[\Hat{D}_\nu b_\nu(t,x),\omega(t,x)].
\label{eq:(2.17)}
\end{equation}
From these expressions, we see that under~Eq.~\eqref{eq:(2.12)}, the flow
equation~\eqref{eq:(2.5)} changes to
\begin{equation}
   \partial_tB_\mu(t,x)=D_\nu G_{\nu\mu}(t,x)
   +\alpha_0D_\mu\Hat{D}_\nu b_\nu(t,x)
   -D_\mu(\partial_t-\alpha_0D_\nu\Hat{D}_\nu)\omega(t,x).
\label{eq:(2.18)}
\end{equation}
This shows that, by choosing $\omega(t,x)$ as a solution of
\begin{equation}
   (\partial_t-\alpha_0D_\nu\Hat{D}_\nu)\omega(t,x)
   =-\delta\alpha_0\Hat{D}_\nu b_\nu(t,x),\qquad
   \omega(t=0,x)=0,
\label{eq:(2.19)}
\end{equation}
the transformed flowed field (that has the same initial value as the original
one) obeys the flow equation~\eqref{eq:(2.5)}
with~$\alpha_0\to\alpha_0+\delta\alpha_0$. Since a gauge-invariant quantity
that does not contain the $t$~derivative is invariant
under~Eq.~\eqref{eq:(2.12)}, this shows that such a gauge-invariant quantity is
independent of~$\alpha_0$. Physical observables are not affected by the gauge
fixing term in~Eq.~\eqref{eq:(2.5)}; the introduction of the gauge-fixing term
is thus a physically allowed modification of the flow equation.

\subsection{Classical perturbative solution to the flow equation}
Now, using Eq.~\eqref{eq:(2.7)} in~Eq.~\eqref{eq:(2.5)}, we have the flow
equation for the quantum field:
\begin{equation}
   \partial_t b_\mu(t,x)
   =\left[\delta_{\mu\nu}\Hat{D}^2+(\alpha_0-1)\Hat{D}_\mu\Hat{D}_\nu\right]
   b_\nu(t,x)
   +2[\Hat{G}_{\mu\nu}(t,x),b_\nu(t,x)]
   +\Hat{R}_\mu(t,x),
\label{eq:(2.20)}
\end{equation}
where
\begin{align}
   \Hat{R}_\mu(t,x)
   &\equiv2[b_\nu(t,x),\Hat{D}_\nu b_\mu(t,x)]
   -[b_\nu(t,x),\Hat{D}_\mu b_\nu(t,x)]
\notag\\
   &\qquad{}
   +(\alpha_0-1)[b_\mu(t,x),\Hat{D}_\nu b_\nu(t,x)]
   +[b_\nu(t,x),[b_\nu(t,x),b_\mu(t,x)]].
\label{eq:(2.21)}
\end{align}
The adjoint actions in these expressions can conveniently be expressed in terms
of matrix multiplication, if one introduces the adjoint representation. We thus
define
\begin{align}
   \Hat{\mathcal{B}}_\mu^{ab}(t,x)
   &\equiv\Hat{B}_\mu^c(t,x)f^{acb},
\label{eq:(2.22)}
\\
   \Hat{\mathcal{D}}_\mu^{ab}
   &\equiv\delta^{ab}\partial_\mu+\Hat{\mathcal{B}}_\mu^{ab}(t,x),
\label{eq:(2.23)}
\\
   \Hat{\mathcal{G}}_{\mu\nu}^{ab}(t,x)
   &\equiv\Hat{G}_{\mu\nu}^c(t,x)f^{acb}.
\label{eq:(2.24)}
\end{align}
With these notations, the flow equation for the quantum field reads
\begin{equation}
   \partial_t b_\mu^a(t,x)
   =\left[\delta_{\mu\nu}\Hat{\mathcal{D}}^2
   +(\alpha_0-1)\Hat{\mathcal{D}}_\mu\Hat{\mathcal{D}}_\nu
   +2\Hat{\mathcal{G}}_{\mu\nu}(t,x)\right]^{ab}
   b_\nu^b(t,x)
   +\Hat{R}_\mu^a(t,x),
\label{eq:(2.25)}
\end{equation}
where
\begin{align}
   \Hat{R}_\mu^a(t,x)
   &=2f^{abc}b_\nu^b(t,x)\Hat{\mathcal{D}}_\nu^{cd}b_\mu^d(t,x)
   -f^{abc}b_\nu^b(t,x)\Hat{\mathcal{D}}_\mu^{cd}b_\nu^d(t,x)
\notag\\
   &\qquad{}
   +(\alpha_0-1)f^{abc}b_\mu^b(t,x)\Hat{\mathcal{D}}_\nu^{cd}b_\nu^d(t,x)
   +f^{abc}f^{cde}b_\nu^b(t,x)b_\nu^d(t,x)b_\mu^e(t,x).
\label{eq:(2.26)}
\end{align}

A formal solution to~Eq.~\eqref{eq:(2.25)} is then given by
\begin{align}
   b_\mu^a(t,x)=\int\mathrm{d}^4y\,\left[
   \Hat{K}_t^{ab}(x,y)_{\mu\nu}a_\nu^b(y)
   +\int_0^t\mathrm{d}s\,\Hat{K}_{t-s}^{ab}(x,y)_{\mu\nu}\Hat{R}_\nu^b(s,y)\right],
\label{eq:(2.27)}
\end{align}
where the heat kernel~$\Hat{K}_t^{ab}(x,y)_{\mu\nu}$ is defined as an object that
satisfies
\begin{align}
   \partial_t\Hat{K}_t^{ab}(x,y)_{\mu\nu}
   &=\left[\delta_{\mu\lambda}\Hat{\mathcal{D}}^2
   +(\alpha_0-1)\Hat{\mathcal{D}}_\mu\Hat{\mathcal{D}}_\lambda
   +2\Hat{\mathcal{G}}_{\mu\lambda}(t,x)\right]^{ac}
   \Hat{K}_t^{cb}(x,y)_{\lambda\nu},
\label{eq:(2.28)}
\\
   \Hat{K}_{t=0}^{ab}(x,y)_{\mu\nu}
   &=\delta^{ab}\delta_{\mu\nu}\delta(x-y).
\label{eq:(2.29)}
\end{align}

The heat kernel defined by~Eqs.~\eqref{eq:(2.28)} and~\eqref{eq:(2.29)} may be
expressed in the form of a time-ordered product containing the flowed
background field. If one is considering a particular situation in which the
background field~$\Hat{A}(x)$ can be assumed to obey the Yang--Mills equation
of motion,
\begin{equation}
   \Hat{D}_\nu\Hat{F}_{\nu\mu}(x)=0,
\label{eq:(2.30)}
\end{equation}
then Eq.~\eqref{eq:(2.7)} implies that the background gauge field does not
flow:
\begin{equation}
   \Hat{B}(t,x)=\Hat{A}(x).
\label{eq:(2.31)}
\end{equation}
Then the heat kernel in the ``Feynman gauge'' $\alpha_0=1$ can be written,
suppressing the gauge and Lorentz indices, in a very compact form:
\begin{align}
   \Hat{K}_t(x,y)
   &=T\exp
   \left\{
   \int_0^t\mathrm{d}s\,
   \left[\Hat{\mathcal{D}}_x^2
   +2\Hat{\mathcal{G}}(s,x)\right]\right\}
   \delta(x-y)
\notag\\
   &=\mathrm{e}^{t\left[\Hat{\mathcal{D}}_x^2
   +2\Hat{\mathcal{F}}(x)\right]}
   \delta(x-y).
\label{eq:(2.32)}
\end{align}
In the last expression, the covariant derivative is defined with respect to
the background gauge field at vanishing flow time, $\Hat{A}_\mu(x)$; we have
also introduced the corresponding field strength in the adjoint representation,
\begin{equation}
   \Hat{\mathcal{F}}_{\mu\nu}^{ab}(x)
   \equiv\Hat{F}_{\mu\nu}^c(x)f^{acb}.
\label{eq:(2.33)}
\end{equation}
In the application to the small flow time expansion in the next section, we can
assume Eq.~\eqref{eq:(2.30)} without loss of generality. We can then use
Eq.~\eqref{eq:(2.32)} for the heat kernel which greatly simplifies the
computation.

\subsection{Tree-level propagator of the flowed gauge field}
So far we have considered the flow equation~\eqref{eq:(2.20)} at the classical
level. The quantum field at vanishing flow time, $a_\nu^b(y)$, contained
in~Eq.~\eqref{eq:(2.27)} is actually the subject of the functional integral
with the Boltzmann weight, specified by the Yang--Mills action
\begin{equation}
   S=\frac{1}{4g_0^2}\int\mathrm{d}^4x\,
   F_{\mu\nu}^a(x)F_{\mu\nu}^a(x)
\label{eq:(2.34)}
\end{equation}
and the gauge-fixing term in the background gauge~\cite{'tHooft:1975vy,%
DeWitt:1980jv,Boulware:1980av,Abbott:1980hw,Ichinose:1981uw}
\begin{equation}
   S_{\text{gauge-fixing}}
   =\frac{\lambda_0}{2g_0^2}\int\mathrm{d}^4x\,
   \Hat{D}_\mu a_\mu^a(x)\Hat{D}_\nu a_\nu^a(x),
\label{eq:(2.35)}
\end{equation}
which also preserves covariance under the background gauge transformation.
Then, in the presence of the background field, the action quadratic in the
quantum field is given by
\begin{align}
   &\left.(S+S_{\text{gauge-fixing}})\right|_{O(a^2)}
\notag\\
   &=-\frac{1}{2g_0^2}\int\mathrm{d}^4x\,
   a_\mu^a(x)
   \left[
   \delta_{\mu\nu}\Hat{\mathcal{D}}^2
   +(\lambda_0-1)\Hat{\mathcal{D}}_\mu\Hat{\mathcal{D}}_\nu
   +2\Hat{\mathcal{F}}_{\mu\nu}(x)
   \right]^{ab}a_\nu^b(x),
\label{eq:(2.36)}
\end{align}
and thus the tree-level propagator in the Feynman gauge~$\lambda_0=1$ is
written as
\begin{equation}
   \left\langle a_\mu^a(x)a_\nu^b(y)\right\rangle_0
    =g_0^2\left(
   \frac{-1}{\Hat{\mathcal{D}}_x^2+2\Hat{\mathcal{F}}(x)}
   \right)_{\mu\nu}^{ab}\delta(x-y).
\label{eq:(2.37)}
\end{equation}

If one can further assume Eq.~\eqref{eq:(2.30)} for the background field, then
the heat kernel (in the Feynman gauge) is given by~Eq.~\eqref{eq:(2.32)}. Then,
from~Eq.~\eqref{eq:(2.27)}, the tree-level propagator of the flowed quantum
field, in the presence of the background field, is given by
\begin{align}
   \left\langle b_\mu^a(t,x)b_\nu^b(s,y)\right\rangle_0
   &=g_0^2\left(
   \mathrm{e}^{t[\Hat{\mathcal{D}}_x^2+2\Hat{\mathcal{F}}(x)]}\,
   \frac{-1}{\Hat{\mathcal{D}}_x^2+2\Hat{\mathcal{F}}(x)}
   \right)_{\mu\rho}^{ac}
   \left(\mathrm{e}^{s[\Hat{\mathcal{D}}_y^2+2\Hat{\mathcal{F}}(y)]}
   \right)_{\nu\rho}^{bc}
   \delta(x-y)
\notag\\
   &=g_0^2\left(
   \mathrm{e}^{(t+s)[\Hat{\mathcal{D}}_x^2+2\Hat{\mathcal{F}}(x)]}\,
   \frac{-1}{\Hat{\mathcal{D}}_x^2+2\Hat{\mathcal{F}}(x)}
   \right)_{\mu\nu}^{ab}\delta(x-y),
\label{eq:(2.38)}
\end{align}
where in the last equality we have noted
\begin{equation}
   \left[\Hat{\mathcal{D}}_y^2+2\Hat{\mathcal{F}}(y)\right]_{\mu\nu}^{ab}
   \delta(x-y)
   =\left[\Hat{\mathcal{D}}_x^2+2\Hat{\mathcal{F}}(x)\right]_{\nu\mu}^{ba}
   \delta(x-y).
\label{eq:(2.39)}
\end{equation}
The above expression~\eqref{eq:(2.38)}, which is manifestly covariant under the
background gauge transformation, will be fully employed in our application to
the small flow time expansion in the next section.

\subsection{Fermion flow}
We can also consider the ``background covariant gauge'' in the flow of fermion
fields~\cite{Luscher:2013cpa}; we adopt the following flow equations:
\begin{align}
   \partial_t\chi(t,x)&=\left\{D^2-\alpha_0[\Hat{D}_\mu b_\mu(t,x)]\right\}
   \chi(t,x),&
   \chi(t=0,x)&=\psi(x),
\label{eq:(2.40)}
\\
   \partial_t\Bar{\chi}(t,x)
   &=\Bar{\chi}(t,x)
   \left\{\overleftarrow{D}^2
   +\alpha_0[\Hat{D}_\mu b_\mu(t,x)]\right\},&
   \Bar{\chi}(t=0,x)&=\Bar{\psi}(x),
\label{eq:(2.41)}
\end{align}
where the covariant derivatives on the fermion fields are defined by
\begin{equation}
   D_\mu=\partial_\mu+B_\mu,\qquad
   \overleftarrow{D}_\mu\equiv\overleftarrow{\partial}_\mu-B_\mu,
\label{eq:(2.42)}
\end{equation}
and
\begin{equation}
   \Hat{D}_\mu=\partial_\mu+\Hat{B}_\mu,\qquad
   \Hat{\overleftarrow{D}}_\mu\equiv\overleftarrow{\partial}_\mu-\Hat{B}_\mu.
\label{eq:(2.43)}
\end{equation}
On the other hand, in these expressions and in what follows,
$[\Hat{D}_\mu b_\mu(t,x)]$ stands for the background covariant derivative on
the quantum gauge field, defined in~Eq.~\eqref{eq:(2.6)}.

One can again see that the gauge parameter~$\alpha_0$ is irrelevant for
gauge-invariant quantities with our gauge-fixing terms
in~Eqs.~\eqref{eq:(2.40)} and~\eqref{eq:(2.41)}. To see this, we again consider
the infinitesimal transformation, Eq.~\eqref{eq:(2.12)}, and
\begin{equation}
   \chi(t,x)\to\left[1-\omega(t,x)\right]\chi(t,x),\qquad
   \Bar{\chi}(t,x)\to\Bar{\chi}(t,x)\left[1+\omega(t,x)\right].
\label{eq:(2.44)}
\end{equation}
Then using Eq.~\eqref{eq:(2.15)}, after some calculation, we see that the flow
equations are changed as
\begin{align}
   &\partial_t\chi(t,x)=\left\{D^2-\alpha_0[\Hat{D}_\mu b_\mu(t,x)]\right\}
   \chi(t,x)
   +(\partial_t-\alpha_0D_\mu\Hat{D}_\mu)\omega(t,x)\chi(t,x),
\label{eq:(2.45)}
\\
   &\partial_t\Bar{\chi}(t,x)
   =\Bar{\chi}(t,x)
   \left\{\overleftarrow{D}^2
   +\alpha_0[\Hat{D}_\mu b_\mu(t,x)]\right\}
   -\Bar{\chi}(t,x)(\partial_t-\alpha_0D_\mu\Hat{D}_\mu)\omega(t,x).
\label{eq:(2.46)}
\end{align}
These show that again by choosing $\omega(t,x)$ as the solution
to~Eq.~\eqref{eq:(2.19)}, we can shift $\alpha_0$ to~$\alpha_0+\delta\alpha_0$.
Gauge-invariant quantities (that do not contain the flow time derivative) are
hence not affected by the gauge-fixing terms in~Eqs.~\eqref{eq:(2.40)}
and~\eqref{eq:(2.41)}.

We also decompose the fermion fields into the background part and the quantum
part as
\begin{align}
   \chi(t,x)&=\Hat{\chi}(t,x)+k(t,x),&
   \Bar{\chi}(t,x)&=\Hat{\Bar{\chi}}(t,x)+\Bar{k}(t,x).
\label{eq:(2.47)}
\\
   \psi(x)&=\Hat{\psi}(x)+p(x),&
   \Bar{\psi}(x)&=\Hat{\Bar{\psi}}(x)+\Bar{p}(x),
\label{eq:(2.48)}
\end{align}
and assume that the background fields themselves are evolved according to
\begin{align}
   &\partial_t\Hat{\chi}(t,x)
   =\Hat{D}^2\Hat{\chi}(t,x),&
   \Hat{\chi}(t=0,x)&=\Hat{\psi}(x),
\label{eq:(2.49)}
\\
   &\partial_t\Hat{\Bar{\chi}}(t,x)
   =\Hat{\Bar{\chi}}(t,x)
   \Hat{\overleftarrow{D}}^2,&
   \Hat{\Bar{\chi}}(t=0,x)&=\Hat{\Bar{\psi}}(x).
\label{eq:(2.50)}
\end{align}
Then, from~Eqs.~\eqref{eq:(2.40)} and~\eqref{eq:(2.41)}, the quantum fields
obey the flow equations
\begin{align}
   \partial_tk(t,x)
   &=\left\{D^2-\alpha_0[\Hat{D}_\mu b_\mu(t,x)]\right\}k(t,x)
\notag\\
   &\qquad{}
   +\left\{
   (1-\alpha_0)[\Hat{D}_\mu b_\mu(t,x)]
   +2b_\mu(t,x)\Hat{D}_\mu+b_\mu(t,x)b_\mu(t,x)
   \right\}
   \Hat{\chi}(t,x),
\notag\\
   &\qquad\qquad\qquad\qquad\qquad\qquad\qquad\qquad\qquad\qquad\qquad
   k(t=0,x)=p(x),
\label{eq:(2.51)}
\\
   \partial_t\Bar{k}(t,x)
   &=\Bar{k}(t,x)
   \left\{\overleftarrow{D}^2
   +\alpha_0[\Hat{D}_\mu b_\mu(t,x)]\right\}
\notag\\
   &\qquad{}
   +\Hat{\Bar{\chi}}(t,x)
   \left\{
   -(1-\alpha_0)[\Hat{D}_\mu b_\mu(t,x)]
   -2\Hat{\overleftarrow{D}}_\mu b_\mu(t,x)
   +b_\mu(t,x)b_\mu(t,x)
   \right\},
\notag\\
   &\qquad\qquad\qquad\qquad\qquad\qquad\qquad\qquad\qquad\qquad\qquad
   \Bar{k}(t=0,x)=\Bar{p}(x).
\label{eq:(2.52)}
\end{align}

If we further assume that the background gauge field fulfills
Eq.~\eqref{eq:(2.30)}, the background gauge field does not evolve
as~Eq.~\eqref{eq:(2.31)} and we can write down relatively simple expressions
for the solution of the fermion flow. The solution to the flow
equations~\eqref{eq:(2.49)} and~\eqref{eq:(2.50)} can be expressed as
\begin{equation}
   \Hat{\chi}(t,x)=\mathrm{e}^{t\Hat{D}^2}\Hat{\psi}(x),\qquad
   \Hat{\Bar{\chi}}(t,x)
   =\Hat{\Bar{\psi}}(x)\mathrm{e}^{t\Hat{\overleftarrow{D}}^2}.
\label{eq:(2.53)}
\end{equation}
Then, the solution to the flow equations~\eqref{eq:(2.51)}
and~\eqref{eq:(2.52)} is given by
\begin{align}
   k(t,x)&=\mathrm{e}^{t\Hat{D}^2}p(x)
   +\int_0^t\mathrm{d}s\,
   \mathrm{e}^{(t-s)\Hat{D}^2}\left[
   2b_\mu(s,x)\Hat{D}_\mu+b_\mu(s,x)b_\mu(s,x)
   \right]
   \left[\mathrm{e}^{s\Hat{D}^2}\Hat{\psi}(x)+k(s,x)\right],
\label{eq:(2.54)}
\\
   \Bar{k}(t,x)&=\Bar{p}(x)\mathrm{e}^{t\Hat{\overleftarrow{D}}^2}
   +\int_0^t\mathrm{d}s\,
   \left[\Hat{\Bar{\psi}}(x)\mathrm{e}^{s\Hat{\overleftarrow{D}}^2}
   +\Bar{k}(s,x)\right]
   \left[-2\Hat{\overleftarrow{D}}_\mu b_\mu(s,x)
   +b_\mu(s,x)b_\mu(s,x)\right]
   \mathrm{e}^{(t-s)\Hat{\overleftarrow{D}}^2},
\label{eq:(2.55)}
\end{align}
where we have adopted the ``Feynman gauge'' $\alpha_0=1$ for simplicity.

The quantum fields at vanishing flow time, $p(x)$ and~$\Bar{p}(x)$, are
subjects of the functional integral with the conventional action,
\begin{align}
   S&=\int\mathrm{d}^Dx\,\Bar{\psi}(x)(\Slash{D}+m_0)\psi(x)
\notag\\
   &=\int\mathrm{d}^Dx\,\left[\Hat{\Bar{\psi}}(x)+\Bar{p}(x)\right]
   \left(\Hat{\Slash{D}}+\Slash{a}+m_0\right)
   \left[\Hat{\psi}(x)+p(x)\right].
\label{eq:(2.56)}
\end{align}
Thus the tree-level propagator of quantum fermion fields, in the presence of
the background gauge field, is given by
\begin{equation}
   \left\langle p(x)\Bar{p}(y)\right\rangle_0
   =\frac{1}{\Hat{\Slash{D}}_x+m_0}\delta(x-y).
\label{eq:(2.57)}
\end{equation}

\section{Application: Small flow time expansion relevant to the energy--momentum tensor}
\label{sec:3}
As noted in~Ref.~\cite{Luscher:2011bx}, any local composite operator of flowed
fields can be expressed as, in the limit of~$t\to0$, an asymptotic series of
local composite operators of fields at vanishing flow time.
In~Ref.~\cite{Suzuki:2013gza}, this \emph{small flow time expansion\/} (with
use of perturbation theory) was exploited to construct a universal formula for
the energy--momentum tensor. This formula with lattice regularization was then
numerically tested in~Ref.~\cite{Asakawa:2013laa} by applying it to the bulk
thermodynamics of quenched QCD. The universal formula can be generalized to
general vector-like gauge theories~\cite{Makino:2014taa} and to various
asymptotically free theories~\cite{Makino:2014sta,Makino:2014cxa,%
Suzuki:2015fka}. In~Refs.~\cite{DelDebbio:2013zaa,Patella:2014dsa},
application of the gradient flow to the lattice energy--momentum tensor is
studied from a somewhat different perspective.

In the present paper, we restrict ourselves to the case of the pure Yang--Mills
theory~\cite{Suzuki:2013gza} and consider the small flow time expansion in
the form,\footnote{In Appendix~\ref{sec:B}, we compute the small flow time
expansion of an operator corresponding to the topological density---another
gauge-invariant dimension-four operator.}
\begin{align}
   &G_{\mu\rho}^a(t,x)G_{\nu\rho}^a(t,x)
\notag\\
   &\stackrel{t\to0}{\sim}
   \left\langle G_{\mu\rho}^a(t,x)G_{\nu\rho}^a(t,x)\right\rangle
   +\zeta_{11}(t)F_{\mu\rho}^a(x)F_{\nu\rho}^a(x)
   +\zeta_{12}(t)\delta_{\mu\nu}F_{\rho\sigma}^a(x)F_{\rho\sigma}^a(x)
   +O(t),
\label{eq:(3.1)}
\end{align}
where the $O(t)$~term is the contribution of composite operators of the mass
dimension being equal to or greater than six. Because of symmetry, only the
above two four-dimensional operators can appear on the right-hand side. The
expansion coefficients can be evaluated in perturbation theory and, to the
one-loop order, we write
\begin{equation}
   \zeta_{11}(t)=1+\zeta_{11}^{(1)}(t)+\dotsb,\qquad
   \zeta_{12}(t)=0+\zeta_{12}^{(1)}(t)+\dotsb.
\label{eq:(3.2)}
\end{equation}
If these coefficients are obtained in the dimensional regularization (with the
spacetime dimension~$D=4-2\epsilon$), then the correctly normalized conserved
energy--momentum tensor (with the vacuum expectation value subtracted) can be
written as~\cite{Suzuki:2013gza,Makino:2014taa}
\begin{align}
   \{T_{\mu\nu}\}_R(x)&=\lim_{t\to0}
   \biggl\{c_1(t)\left[G_{\mu\rho}^a(t,x)G_{\nu\rho}^a(t,x)
   -\frac{1}{4}\delta_{\mu\nu}G_{\rho\sigma}^a(t,x)G_{\rho\sigma}^a(t,x)\right]
\notag\\
   &\qquad\qquad{}
   +c_2(t)
   \left[\delta_{\mu\nu}G_{\rho\sigma}^a(t,x)G_{\rho\sigma}^a(t,x)
   -\left\langle\delta_{\mu\nu}G_{\rho\sigma}^a(t,x)G_{\rho\sigma}^a(t,x)
   \right\rangle\right]\biggr\},
\label{eq:(3.3)}
\end{align}
where
\begin{equation}
   c_1(t)=\frac{1}{g_0^2}\left[1-\zeta_{11}^{(1)}(t)\right],\qquad
   c_2(t)=\frac{1}{g_0^2}
   \left[-\frac{1}{2}\epsilon\zeta_{12}^{(1)}(t)\right].
\label{eq:(3.4)}
\end{equation}

Since bare composite operators of the flowed gauge field are automatically
renormalized operators~\cite{Luscher:2011bx}, the formula~\eqref{eq:(3.3)}
should hold irrespective of an adopted regularization;\footnote{The
coefficients $c_1(t)$ and~$c_2(t)$ in~Eq.~\eqref{eq:(3.4)} become finite
for~$\epsilon\to0$ when expressed in terms of renormalized quantities; see
below.} in this sense the formula is universal. In particular, it should hold
with lattice regularization with which the construction of a
correctly normalized conserved energy--momentum tensor is not straightforward.
It is thus of great interest to compute the expansion coefficients
in~Eq.~\eqref{eq:(3.1)}. As we will see below, the background field method we
have developed provides a very efficient non-diagrammatic computational method
of the expansion coefficients (at least in the one-loop level).

Now, to determine the expansion coefficients $\zeta_{11}(t)$
and~$\zeta_{12}(t)$ in~Eq.~\eqref{eq:(3.1)}, we consider 1PI diagrams
containing the composite operators~$G_{\mu\rho}^a(t,x)G_{\nu\rho}^a(t,x)$
and~$F_{\mu\rho}^a(x)F_{\nu\rho}^a(x)$ with external lines of the background
gauge field~$\Hat{B}_\mu(t,x)$ only (i.e., no external line of the quantum
field). In the tree level, since the flow time evolution is purely classical,
\begin{align}
   \left\langle
   G_{\mu\rho}^a(t,x)G_{\nu\rho}^a(t,x)
   \right\rangle_{\text{1PI}}
   &\stackrel{t\to0}{\sim}\Hat{F}_{\mu\rho}^a(x)\Hat{F}_{\nu\rho}^a(x)+O(t),
\label{eq:(3.5)}
\\
   \left\langle
   F_{\mu\rho}^a(x)F_{\nu\rho}^a(x)
   \right\rangle_{\text{1PI}}
   &=\Hat{F}_{\mu\rho}^a(x)\Hat{F}_{\nu\rho}^a(x).
\label{eq:(3.6)}
\end{align}
Comparing these two relations, we find
\begin{equation}
   G_{\mu\rho}^a(t,x)G_{\nu\rho}^a(t,x)
   \stackrel{t\to0}{\sim}
   F_{\mu\rho}^a(x)F_{\nu\rho}^a(x)+O(t).
\label{eq:(3.7)}
\end{equation}
This gives the tree-level contributions in~Eq.~\eqref{eq:(3.2)}.

Next, we consider one-loop 1PI diagrams containing the composite operators
with external lines of the background gauge field. Such 1PI diagrams can be
obtained, by taking the contraction of quantum fields in the expansion of the
composite operators by the propagator in the presence of the background field.
The expansion of the composite operator~$G_{\mu\rho}^a(t,x)G_{\nu\rho}^a(t,x)$ in
the quadratic order yields
\begin{align}
   &\left.G_{\mu\rho}^a(t,x)G_{\nu\rho}^a(t,x)\right|_{O(b^2)}
\notag\\
   &=(\delta_{\mu\alpha}\delta_{\nu\delta}\delta_{\beta\gamma}
   -\delta_{\mu\alpha}\delta_{\nu\gamma}\delta_{\beta\delta}
   -\delta_{\mu\beta}\delta_{\nu\delta}\delta_{\alpha\gamma}
   +\delta_{\mu\beta}\delta_{\nu\gamma}\delta_{\alpha\delta})
   \left[\Hat{\mathcal{D}}_\alpha b_\beta(t,x)\right]^a
   \left[\Hat{\mathcal{D}}_\delta b_\gamma(t,x)\right]^a
\notag\\
   &\qquad{}
   -b_\nu(t,x)\Hat{\mathcal{F}}_{\mu\rho}(x)b_\rho(t,x)
   -b_\mu(t,x)\Hat{\mathcal{F}}_{\nu\rho}(x)b_\rho(t,x).
\label{eq:(3.8)}
\end{align}
At this stage we note that to read off the expansion coefficients
in~Eq.~\eqref{eq:(3.1)} as~Eq.~\eqref{eq:(3.7)}, we can assume that the
background field satisfies the Yang--Mills equation of motion,
Eq~\eqref{eq:(2.30)}, because $\Hat{F}_{\mu\rho}^a(x)\Hat{F}_{\nu\rho}^a(x)$ does
not vanish under the equation of motion. Then the background field does not
flow~$\Hat{B}(t,x)=\Hat{A}(x)$ and we can use the simple
expression~\eqref{eq:(2.38)} for the propagator. The contraction then yields
\begin{align}
   &\left\langle
   \left.G_{\mu\rho}^a(t,x)G_{\nu\rho}^a(t,x)\right|_{O(b^2)}
   \right\rangle_{\text{1PI}}
\notag\\
   &=g_0^2(\delta_{\mu\alpha}\delta_{\nu\delta}\delta_{\beta\gamma}
   -\delta_{\mu\alpha}\delta_{\nu\gamma}\delta_{\beta\delta}
   -\delta_{\mu\beta}\delta_{\nu\delta}\delta_{\alpha\gamma}
   +\delta_{\mu\beta}\delta_{\nu\gamma}\delta_{\alpha\delta})
   \Hat{\mathcal{D}}_\alpha^{ab}
   \left(\mathrm{e}^{2t\Hat{\Delta}}\frac{1}{\Hat{\Delta}}
   \right)_{\beta\gamma}^{bc}
   \Hat{\mathcal{D}}_\delta^{ca}\left.\delta(x-y)\right|_{y=x}
\notag\\
   &\qquad{}
   +g_0^2\Hat{\mathcal{F}}_{\mu\rho}^{ab}(x)
   \left(\mathrm{e}^{2t\Hat{\Delta}}\frac{1}{\Hat{\Delta}}
   \right)_{\rho\nu}^{ba}\left.\delta(x-y)\right|_{y=x}
   +g_0^2\Hat{\mathcal{F}}_{\nu\rho}^{ab}(x)
   \left(\mathrm{e}^{2t\Hat{\Delta}}\frac{1}{\Hat{\Delta}}
   \right)_{\rho\mu}^{ba}\left.\delta(x-y)\right|_{y=x},
\label{eq:(3.9)}
\end{align}
where
\begin{equation}
   \Hat{\Delta}\equiv\hat{\mathcal{D}}^2+2\Hat{\mathcal{F}},
\label{eq:(3.10)}
\end{equation}
and we have noted
\begin{equation}
   \Hat{\mathcal{D}}_{y\delta}^{ac}\delta(x-y)
   =-\Hat{\mathcal{D}}_{x\delta}^{ca}\delta(x-y).
\label{eq:(3.11)}
\end{equation}

However, as explained in~Refs.~\cite{Makino:2014taaa} (see also
Ref.~\cite{Makino:2014sta}), instead of the 1PI function~\eqref{eq:(3.9)}
itself, it is much convenient to consider the difference
\begin{equation}
   \left\langle
   \left.G_{\mu\rho}^a(t,x)G_{\nu\rho}^a(t,x)\right|_{O(b^2)}
   -\left.F_{\mu\rho}^a(x)F_{\nu\rho}^a(x)\right|_{O(a^2)}
   \right\rangle_{\text{1PI}},
\label{eq:(3.12)}
\end{equation}
because possible infrared divergences are cancelled out in this combination.
The one-loop 1PI function, which contains $F_{\mu\rho}^a(x)F_{\nu\rho}^a(x)$ is
given by simply setting~$t=0$ in~Eq.~\eqref{eq:(3.9)}. Then the difference can
be expressed as an integral over an auxiliary variable~$\xi$ as,
\begin{align}
   &\left\langle
   \left.G_{\mu\rho}^a(t,x)G_{\nu\rho}^a(t,x)\right|_{O(b^2)}
   -\left.F_{\mu\rho}^a(x)F_{\nu\rho}^a(x)\right|_{O(a^2)}
   \right\rangle_{\text{1PI}}
\notag\\
   &=2g_0^2\int_0^t\mathrm{d}\xi\,\biggl[
   (\delta_{\mu\alpha}\delta_{\nu\delta}\delta_{\beta\gamma}
   -\delta_{\mu\alpha}\delta_{\nu\gamma}\delta_{\beta\delta}
   -\delta_{\mu\beta}\delta_{\nu\delta}\delta_{\alpha\gamma}
   +\delta_{\mu\beta}\delta_{\nu\gamma}\delta_{\alpha\delta})
   \Hat{\mathcal{D}}_\alpha^{ab}
   \left(\mathrm{e}^{2\xi\Hat{\Delta}}\right)_{\beta\gamma}^{bc}
   \Hat{\mathcal{D}}_\delta^{ca}
\notag\\
   &\qquad\qquad\qquad\qquad{}
   +\Hat{\mathcal{F}}_{\mu\rho}^{ab}(x)
   \left(\mathrm{e}^{2\xi\Hat{\Delta}}
   \right)_{\rho\nu}^{ba}
   +\Hat{\mathcal{F}}_{\nu\rho}^{ab}(x)
   \left(\mathrm{e}^{2\xi\Hat{\Delta}}
   \right)_{\rho\mu}^{ba}\biggr]\left.\delta(x-y)\right|_{y=x}.
\label{eq:(3.13)}
\end{align}
In this expression, it is obvious that there is no infrared divergence, because
derivative operators appear only in positive powers.

We then set
\begin{equation}
   \delta(x-y)=\int\frac{\mathrm{d}^Dp}{(2\pi)^D}\,\mathrm{e}^{ipx}
   \mathrm{e}^{-ipy},
\label{eq:(3.14)}
\end{equation}
and moves the plain wave~$\mathrm{e}^{ipx}$ the most left-hand side, by noting
\begin{equation}
   \Hat{\mathcal{D}}_\mu\mathrm{e}^{ipx}
   =\mathrm{e}^{ipx}(ip_\mu+\Hat{\mathcal{D}}_\mu),
\label{eq:(3.15)}
\end{equation}
as usual in the calculation of anomalies in the path
integral~\cite{Fujikawa:1979ay,Fujikawa:1980eg,Fujikawa:1980vr,%
Fujikawa:1983az,Fujikawa:1993xv}. After the rescaling of the integration
variables, $p_\mu\to p_\mu/\sqrt{\xi}$, we have
\begin{align}
   &\left\langle
   \left.G_{\mu\rho}^a(t,x)G_{\nu\rho}^a(t,x)\right|_{O(b^2)}
   -\left.F_{\mu\rho}^a(x)F_{\nu\rho}^a(x)\right|_{O(a^2)}
   \right\rangle_{\text{1PI}}
\notag\\
   &=2g_0^2\int_0^t\mathrm{d}\xi\,\xi^{-D/2}
   \int\frac{\mathrm{d}^Dp}{(2\pi)^D}\,\mathrm{e}^{-2p^2}
\notag\\
   &\qquad{}\times
   \tr\biggl[
   (\delta_{\mu\alpha}\delta_{\nu\delta}\delta_{\beta\gamma}
   -\delta_{\mu\alpha}\delta_{\nu\gamma}\delta_{\beta\delta}
   -\delta_{\mu\beta}\delta_{\nu\delta}\delta_{\alpha\gamma}
   +\delta_{\mu\beta}\delta_{\nu\gamma}\delta_{\alpha\delta})
\notag\\
   &\qquad\qquad\qquad{}\times
   \xi^{-1}
   \left(ip_\alpha+\sqrt{\xi}\Hat{\mathcal{D}}_\alpha\right)
   \left(\mathrm{e}^{4i\sqrt{\xi}p\cdot\Hat{\mathcal{D}}+2\xi\Hat{\Delta}}
   \right)_{\beta\gamma}
   \left(ip_\delta+\sqrt{\xi}\Hat{\mathcal{D}}_\delta\right)
\notag\\
   &\qquad\qquad\qquad\qquad{}
   +\Hat{\mathcal{F}}_{\mu\rho}(x)
   \left(\mathrm{e}^{4i\sqrt{\xi}p\cdot\Hat{\mathcal{D}}+2\xi\Hat{\Delta}}
   \right)_{\rho\nu}
   +\Hat{\mathcal{F}}_{\nu\rho}(x)
   \left(\mathrm{e}^{4i\sqrt{\xi}p\cdot\Hat{\mathcal{D}}+2\xi\Hat{\Delta}}
   \right)_{\rho\mu}
   \biggr],
\label{eq:(3.16)}
\end{align}
where the trace~$\tr$ is for the gauge indices. It is interesting to note that
all the information of relevant one-loop 1PI diagrams is contained in this
single compact expression; in a conventional calculational
scheme~\cite{Suzuki:2013gza}, on the other hand, one has to compute at least 12
1PI diagrams to obtain the expansion coefficients in~Eq.~\eqref{eq:(3.1)}.

Now, since we are interested in the small flow time limit~$t\to0$
of~Eq.~\eqref{eq:(3.16)} and since~$\xi\in[0,t]$, we may expand the integrand
with respect to~$\xi$. For $t\to0$, only terms to~$O(\xi^{-D/2+1})$ under the
integral can give rise to non-vanishing contributions for~$D\to4$. The
expansion of the
combination~$\mathrm{e}^{4i\sqrt{\xi}p\cdot\Hat{\mathcal{D}}+2\xi\Hat{\Delta}}$
to~$O(\xi^2)$ is given in~Appendix~\ref{sec:A}. Although the remaining
algebraic calculation after the Gaussian integration over~$p$, by noting
\begin{equation}
   [\Hat{\mathcal{D}}_\rho,\Hat{\mathcal{D}}_\sigma]
   =\Hat{\mathcal{F}}_{\rho\sigma}
\label{eq:(3.17)}
\end{equation}
is somewhat lengthy, it is rather straightforward. In this calculation, it is
quite helpful to note that the final expression for~Eq.~\eqref{eq:(3.16)} must
be symmetric under~$\mu\leftrightarrow\nu$ by definition; we may thus simply
discard any terms anti-symmetric under~$\mu\leftrightarrow\nu$. In this way, we
finally arrive at
\begin{align}
   &\left\langle
   \left.G_{\mu\rho}^a(t,x)G_{\nu\rho}^a(t,x)\right|_{O(b^2)}
   -\left.F_{\mu\rho}^a(x)F_{\nu\rho}^a(x)\right|_{O(a^2)}
   \right\rangle_{\text{1PI}}
\notag\\
   &\stackrel{t\to0}{\sim}
   \frac{g_0^2}{(4\pi)^2}\dim(G)\frac{3}{8t^2}\delta_{\mu\nu}
\notag\\
   &\qquad{}
   +\frac{g_0^2}{(4\pi)^2}
   \left[
   \frac{11}{3}\epsilon(t)^{-1}+\frac{7}{3}
   \right]\tr\left[\Hat{\mathcal{F}}(x)^2\right]_{\mu\nu}
\notag\\
   &\qquad\qquad{}
   +\frac{g_0^2}{(4\pi)^2}\left[
   -\frac{11}{12}\epsilon(t)^{-1}-\frac{1}{6}
   \right]\delta_{\mu\nu}\tr\left[\Hat{\mathcal{F}}(x)^2\right]_{\rho\rho}
   +O(t),
\label{eq:(3.18)}
\end{align}
where
\begin{equation}
   \epsilon(t)^{-1}\equiv\frac{1}{\epsilon}+\ln(8\pi t).
\label{eq:(3.19)}
\end{equation}
Since
\begin{equation}
   \tr\left[\Hat{\mathcal{F}}(x)^2\right]_{\mu\nu}
   =\Hat{\mathcal{F}}_{\mu\rho}^{ab}(x)\Hat{\mathcal{F}}_{\rho\nu}^{ba}(x)
   =f^{acb}f^{bda}
   \Hat{F}_{\mu\rho}^c(x)\Hat{F}_{\rho\nu}^d(x)
   =C_2(G)\Hat{F}_{\mu\rho}^a(x)\Hat{F}_{\nu\rho}^a(x),
\label{eq:(3.20)}
\end{equation}
recalling the tree-level relations~\eqref{eq:(3.5)} and~\eqref{eq:(3.6)},
Eq.~\eqref{eq:(3.18)} shows that the small flow time expansion to the one-loop
order is given by
\begin{align}
   &G_{\mu\rho}^a(t,x)G_{\nu\rho}^a(t,x)
\notag\\
   &\stackrel{t\to0}{\sim}
   \frac{g_0^2}{(4\pi)^2}\dim(G)\frac{3}{8t^2}\delta_{\mu\nu}
\notag\\
   &\qquad{}
   +\left\{
   1+\frac{g_0^2}{(4\pi)^2}C_2(G)
   \left[\frac{11}{3}\epsilon(t)^{-1}+\frac{7}{3}\right]
   \right\}
   F_{\mu\rho}^a(x)F_{\nu\rho}^a(x)
\notag\\
   &\qquad\qquad{}
   +\frac{g_0^2}{(4\pi)^2}C_2(G)\left[
   -\frac{11}{12}\epsilon(t)^{-1}-\frac{1}{6}
   \right]
   \delta_{\mu\nu}F_{\rho\sigma}^a(x)F_{\rho\sigma}^a(x)
   +O(t).
\label{eq:(3.21)}
\end{align}
From this, $\zeta_{11}^{(1)}$ and~$\zeta_{12}^{(1)}$ in~Eq.~\eqref{eq:(3.2)} are
given by
\begin{align}
   \zeta_{11}^{(1)}(t)
   &=\frac{g_0^2}{(4\pi)^2}C_2(G)
   \left[\frac{11}{3}\epsilon(t)^{-1}+\frac{7}{3}\right],
\label{eq:(3.22)}
\\
   \zeta_{12}^{(1)}(t)
   &=\frac{g_0^2}{(4\pi)^2}C_2(G)
   \left[-\frac{11}{12}\epsilon(t)^{-1}-\frac{1}{6}\right],
\label{eq:(3.23)}
\end{align}
and then Eq.~\eqref{eq:(3.4)} gives $c_1(t)$ and~$c_2(t)$
in~Eq.~\eqref{eq:(3.3)}. In terms of the renormalized gauge coupling~$g$ in the
$\text{MS}$ scheme,
\begin{equation}
   \frac{1}{g_0^2}=\frac{1}{g^2}
   +b_0\left(\frac{1}{\epsilon}-\ln\mu^2\right),
\label{eq:(3.24)}
\end{equation}
where $b_0$ is the one-loop coefficient in the beta function,
\begin{equation}
   b_0=\frac{1}{(4\pi)^2}C_2(G)\frac{11}{3},
\label{eq:(3.25)}
\end{equation}
we have
\begin{align}
   c_1(t)&=\frac{1}{g^2}-b_0\ln(8\pi\mu^2t)
   -\frac{1}{(4\pi)^2}C_2(G)\frac{7}{3},
\label{eq:(3.26)}
\\
   c_2(t)&=\frac{1}{8}b_0.
\label{eq:(3.27)}
\end{align}

The above non-diagrammatic one-loop computation of coefficients $c_1(t)$
and~$c_2(t)$ is much simpler and quicker than the diagrammatic calculation
carried out in~Ref.~\cite{Suzuki:2013gza}. Unfortunately, the results of the
above calculation do not coincide with the results
in~Ref.~\cite{Suzuki:2013gza}, revealing that there are errors in the one-loop
diagrammatic calculation in~Ref.~\cite{Suzuki:2013gza}.\footnote{In particular,
Eq.~(4.30) and~(4.31) of~Ref.~\cite{Suzuki:2013gza} should be
\begin{align}
   c_1&=\ln\sqrt{\pi}+\frac{7}{22}
   \simeq0.890547,
\\
   c_2&=\ln\sqrt{\pi}-\frac{7}{44}+\frac{b_1}{2b_0^2}
   \simeq0.834762.
\end{align}
Equations~(4.32) and~(4.33) of~Ref.~\cite{Makino:2014taa} should be replaced by
Eqs.~\eqref{eq:(3.22)} and~\eqref{eq:(3.23)}, respectively, and
consequently, Eq.~(4.72) of~Ref.~\cite{Makino:2014taa} should be
\begin{equation}
   c_1(t)
   =\frac{1}{\Bar{g}(1/\sqrt{8t})^2}-b_0\ln\pi
   -\frac{1}{(4\pi)^2}
   \left[\frac{7}{3}C_2(G)-\frac{3}{2}T(R)N_{\text{f}}\right],
\end{equation}
where $\Bar{g}(1/\sqrt{8t})$ is the running gauge coupling in the $\text{MS}$
scheme at the renormalization scale~$\mu=1/\sqrt{8t}$.
The expressions just below Eqs.~(5) and~(6) of~Ref.~\cite{Asakawa:2013laa}
should be
\begin{align}
   \Bar{s}_1&=\frac{7}{22}+\frac{1}{2}\gamma_E-\ln2
   \simeq-0.0863575299274,
\\
   \Bar{s}_2&=\frac{21}{44}-\frac{b_1}{2b_0^2}=\frac{27}{484}
   \simeq0.0557851239669.
\end{align}
The erratum for~Ref.~\cite{Asakawa:2013laa} will appear soon.}

\section{Application: Small flow time expansion of the axial-vector current}
\label{sec:4}
As another application of the present formulation, we consider the small flow
time expansion of the axial-vector current of the flowed fermion
fields~\cite{Endo:2015iea}:
\begin{equation}
   \Bar{\chi}(t,x)\gamma_\mu\gamma_5t^A\chi(t,x)\stackrel{t\to0}{\sim}
   \left[1+\xi^{(1)}(t)\right]
   \Bar{\psi}(x)\gamma_\mu\gamma_5t^A\psi(x)+O(t),
\label{eq:(4.1)}
\end{equation}
where $t^A$ is the generator of the flavor symmetry group and $\xi^{(1)}(t)$ is
the expansion coefficient at the one-loop level. Because of symmetry, only the
axial-vector current at vanishing flow time can appear as the leading $O(t^0)$
term in the right-hand side. To find $\xi^{(1)}(t)$, we set the background
gauge field to zero and consider one-loop 1PI diagrams containing the composite
operator~$\Bar{\chi}(t,x)\gamma_\mu\gamma_5t^A\chi(t,x)$ with external lines of
the background fermion fields, $\Hat{\psi}(t,x)$ and~$\Hat{\Bar{\psi}}(t,x)$
(no external line of the quantum fields).

As Eqs.~\eqref{eq:(3.5)} and~\eqref{eq:(3.6)}, at the tree level,
\begin{align}
   \left\langle\Bar{\chi}(t,x)\gamma_\mu\gamma_5t^A\chi(t,x)
   \right\rangle_{\text{1PI}}
   &\stackrel{t\to0}{\sim}
   \Hat{\Bar{\psi}}(x)\gamma_\mu\gamma_5t^A\Hat{\psi}(x)+O(t),
\label{eq:(4.2)}
\\
   \left\langle\Bar{\psi}(x)\gamma_\mu\gamma_5t^A\psi(x)
   \right\rangle_{\text{1PI}}
   &=\Hat{\Bar{\psi}}(x)\gamma_\mu\gamma_5t^A\Hat{\psi}(x).
\label{eq:(4.3)}
\end{align}

For the one-loop level, again as Eq.~\eqref{eq:(3.12)}, it is convenient to
consider the difference:
\begin{equation}
   \left\langle
   \Bar{\chi}(t,x)\gamma_\mu\gamma_5t^A\chi(t,x)
   -\Bar{\psi}(x)\gamma_\mu\gamma_5t^A\psi(x)
   \right\rangle_{\text{1PI}},
\label{eq:(4.4)}
\end{equation}
because this combination is free from infrared divergences. For the first term
of~Eq.~\eqref{eq:(4.4)}, by using the decomposition~\eqref{eq:(2.47)}, we have
\begin{align}
   &\Bar{\chi}(t,x)\gamma_\mu\gamma_5t^A\chi(t,x)
\notag\\
   &=\Hat{\Bar{\chi}}(t,x)\gamma_\mu\gamma_5t^A\Hat{\chi}(t,x)
   +\Hat{\Bar{\chi}}(t,x)\gamma_\mu\gamma_5t^Ak(t,x)
   +\Bar{k}(t,x)\gamma_\mu\gamma_5t^A\Hat{\chi}(t,x)
   +\Bar{k}(t,x)\gamma_\mu\gamma_5t^Ak(t,x).
\label{eq:(4.5)}
\end{align}
We can then use Eqs.~\eqref{eq:(2.54)} and~\eqref{eq:(2.55)} to express the
quantum flowed fields, $k(t,x)$ and~$\Bar{k}(t,x)$, in terms of fermion fields
at vanishing flow time. Equation~\eqref{eq:(4.1)} shows that to find the
coefficient~$\xi^{(1)}(t)$, we may set the background fermion fields to be
constant which makes the calculation quite easy. Then, as terms which
contribute to one-loop 1PI diagrams with external lines of background fermion
fields, we have
\begin{align}
   &k(t,x)=\mathrm{e}^{t\partial^2}p(x)
   +\int_0^t\mathrm{d}s\,\mathrm{e}^{(t-s)\partial^2}b_\mu(s,x)b_\mu(s,x)
   \Hat\psi
   +\int_0^t\mathrm{d}s\,\mathrm{e}^{(t-s)\partial^2}2b_\mu(s,x)\partial_\mu
   \mathrm{e}^{s\partial^2}p(x),
\label{eq:(4.6)}
\\
   &\Bar{k}(t,x)
\notag\\
   &=\Bar{p}(x)\mathrm{e}^{t\overleftarrow{\partial}^2}
   +\int_0^t\mathrm{d}s\,
   \Hat{\Bar{\psi}}
   b_\mu(s,x)b_\mu(s,x)
   \mathrm{e}^{(t-s)\Hat{\overleftarrow{\partial}}^2}
   +\int_0^t\mathrm{d}s\,
   \Bar{p}(x)\mathrm{e}^{s\overleftarrow{\partial}^2}
   (-2)\Hat{\overleftarrow{\partial}}_\mu b_\mu(s,x)
   \mathrm{e}^{(t-s)\Hat{\overleftarrow{\partial}}^2}.
\label{eq:(4.7)}
\end{align}

In Eqs.~\eqref{eq:(4.6)} and~\eqref{eq:(4.7)}, the quantum fields at
vanishing flow time, $p(x)$ and~$\Bar{p}(x)$, are subject to the functional
integral with the action~\eqref{eq:(2.56)}. Through the interaction terms
in~Eq.~\eqref{eq:(2.56)}, $p(x)$ and~$\Bar{p}(x)$ become the background
fields, $\Hat{\psi}$ and~$\Hat{\Bar{\psi}}$. Considering the contraction by
the propagator~\eqref{eq:(2.57)} in
$\langle p(x)(-1)\int\mathrm{d}^Dy\,\Bar{p}(y)\Slash{a}(y)\Hat{\psi}\rangle$
and
$\langle(-1)\int\mathrm{d}^Dy\,\Hat{\Bar{\psi}}
\Slash{a}(y)p(y)\Bar{p}(x)\rangle$,\footnote{Note that we are now setting the
background gauge field to zero.} this effect of interaction vertices can
effectively be represented by the substitutions,
\begin{align}
   p(x)&\to-\int\mathrm{d}^Dy\,\frac{1}{\Slash{\partial}_x+m_0}\delta(x-y)
   \Slash{a}(y)\Hat{\psi},
\label{eq:(4.8)}
\\
   \Bar{p}(x)&
   \to-\int\mathrm{d}^Dy\,\Hat{\Bar{\psi}}\Slash{a}(y)
   \frac{1}{\Slash{\partial}_y+m_0}\delta(y-x).
\label{eq:(4.9)}
\end{align}
Note that these substitutions accompany a gauge interaction vertex coming from
the action~\eqref{eq:(2.56)} because we are considering 1PI diagrams. Then the
contraction of the quantum gauge fields in the expectation value
of~Eq.~\eqref{eq:(4.5)} by the propagator~\eqref{eq:(2.38)} is very simple. In
this way, we have the one-loop expression for the first term
of~Eq.~\eqref{eq:(4.4)}. Then, by simply setting~$t=0$ in that expression, we
have the one-loop expression for the second term of~Eq.~\eqref{eq:(4.4)}. The
resulting difference is free from infrared divergences and under the
dimensional regularization, it is simple to obtain at the one-loop level
\begin{align}
   &\left\langle
   \Bar{\chi}(t,x)\gamma_\mu\gamma_5t^A\chi(t,x)
   -\Bar{\psi}(x)\gamma_\mu\gamma_5t^A\psi(x)
   \right\rangle_{\text{1PI}},
\notag\\
   &\stackrel{t\to0}{\sim}
   \frac{g_0^2}{(4\pi)^2}C_2(R)(-3)
   \left[\epsilon(t)^{-1}+\frac{7}{6}\right]
   \Hat{\Bar{\psi}}(x)\gamma_\mu\gamma_5t^A\Hat{\psi}(x)+O(g_0^4).
\label{eq:(4.10)}
\end{align}
Because of the tree-level relations~\eqref{eq:(4.2)} and~\eqref{eq:(4.3)},
Eq.~\eqref{eq:(4.10)} shows that
\begin{equation}
   \Bar{\chi}(t,x)\gamma_\mu\gamma_5t^A\chi(t,x)
   \stackrel{t\to0}{\sim}
   \left\{1+\frac{g_0^2}{(4\pi)^2}C_2(R)(-3)
   \left[\epsilon(t)^{-1}+\frac{7}{6}\right]
   \right\}\Bar{\psi}(x)\gamma_\mu\gamma_5t^A\psi(x),
\label{eq:(4.11)}
\end{equation}
which coincides with the result in~Ref.~\cite{Endo:2015iea}.

As discussed in~Ref.~\cite{Endo:2015iea}, Eq.~\eqref{eq:(4.11)} shows that the
correctly normalized axial-vector current can be expressed as
\begin{equation}
   j_{5\mu}^A(x)
   =\lim_{t\to0}\left\{1+
   \frac{\Bar{g}(1/\sqrt{8t})^2}{(4\pi)^2}C_2(R)
   \left[-\frac{1}{2}+\ln(432)\right]\right\}
   \mathring{\Bar{\chi}}(t,x)\gamma_\mu\gamma_5t^A\mathring{\chi}(t,x),
\label{eq:(4.12)}
\end{equation}
where~\cite{Makino:2014taa}
\begin{align}
   \mathring{\chi}(t,x)
   &=\sqrt{\frac{-2\dim(R)N_f}
   {(4\pi)^2t^2
   \left\langle\Bar{\chi}(t,x)\overleftrightarrow{\Slash{D}}\chi(t,x)
   \right\rangle}}
   \,\chi(t,x),
\label{eq:(4.13)}
\\
   \mathring{\Bar{\chi}}(t,x)
   &=\sqrt{\frac{-2\dim(R)N_f}
   {(4\pi)^2t^2
   \left\langle\Bar{\chi}(t,x)\overleftrightarrow{\Slash{D}}\chi(t,x)
   \right\rangle}}
   \,\Bar{\chi}(t,x),
\label{eq:(4.14)}
\end{align}
($N_f$ is the number of flavors) and
\begin{equation}
   \overleftrightarrow{D}_\mu\equiv D_\mu-\overleftarrow{D}_\mu.
\label{eq:(4.15)}
\end{equation}

\section{Conclusion}
\label{sec:5}
In the present paper, we have developed a background field method (or a
background gauge covariant gauge fixing, more appropriately) for the gradient
flow equations. This formulation allows a manifestly background gauge covariant
perturbative expansion of the flow equations. We illustrated the power of the
method by applying it to the one-loop calculation of expansion coefficients in
the small flow time expansion relevant to the energy--momentum tensor. This new
simple computational scheme revealed that there were errors in the old
diagrammatic calculation in~Ref.~\cite{Suzuki:2013gza} (the errors have been
identified and corrected~\cite{Suzuki:2013gzaa,Makino:2014taaa}).

Since our method provides a greatly simplified computational scheme for known
one-loop computations, we can expect that it can also be useful in more
complicated situations, such as the two-loop computation of the expansion
coefficients. We hope to come back to possible further applications of the
present formulation in the near future.

\section*{Acknowledgments}
The author would like to thank Kazuo Fujikawa and Hiroki Makino for helpful
discussions.
The work of H.~S. is supported in part by Grant-in-Aid for Scientific
Research~23540330.

\appendix
\section{Expansion of~$\mathrm{e}^{4i\sqrt{\xi}p\cdot\Hat{\mathcal{D}}+2\xi\Hat{\Delta}}$}
\label{sec:A}
A straightforward expansion yields
\begin{align}
   &\left(\mathrm{e}^{4i\sqrt{\xi}p\cdot\Hat{\mathcal{D}}+2\xi\Hat{\Delta}}
   \right)_{\mu\nu}
\notag\\
   &=\delta_{\mu\nu}
   +4ip\cdot\Hat{\mathcal{D}}\delta_{\mu\nu}\xi^{1/2}
   +\left[2\Hat{\Delta}_{\mu\nu}
   -8(p\cdot\Hat{\mathcal{D}})^2\delta_{\mu\nu}\right]\xi
\notag\\
   &\qquad{}
   +\left[
   4i\Hat{\Delta}_{\mu\nu}p\cdot\Hat{\mathcal{D}}
   +4ip\cdot\Hat{\mathcal{D}}\Hat{\Delta}_{\mu\nu}
   -\frac{32}{3}i(p\cdot\Hat{\mathcal{D}})^3\delta_{\mu\nu}
   \right]\xi^{3/2}
\notag\\
   &\qquad{}
   +\left[
   2\Hat{\Delta}^2_{\mu\nu}
   -\frac{16}{3}\Hat{\Delta}_{\mu\nu}(p\cdot\Hat{\mathcal{D}})^2
   -\frac{16}{3}p\cdot\Hat{\mathcal{D}}\Hat{\Delta}_{\mu\nu}p\cdot\Hat{\mathcal{D}}
   -\frac{16}{3}(p\cdot\Hat{\mathcal{D}})^2\Hat{\Delta}_{\mu\nu}
   +\frac{32}{3}(p\cdot\Hat{\mathcal{D}})^4\delta_{\mu\nu}
   \right]\xi^2
\notag\\
   &\qquad{}+O(\xi^{5/2}),
\label{eq:(A1)}
\end{align}
where
\begin{align}
   \Hat{\Delta}_{\mu\nu}
   &=\Hat{\mathcal{D}}^2\delta_{\mu\nu}+2\Hat{\mathcal{F}}_{\mu\nu},
\label{eq:(A2)}
\\
   \Hat{\Delta}^2_{\mu\nu}
   &=\Hat{\mathcal{D}}^2\Hat{\mathcal{D}}^2\delta_{\mu\nu}
   +2\Hat{\mathcal{D}}^2\Hat{\mathcal{F}}_{\mu\nu}
   +2\Hat{\mathcal{F}}_{\mu\nu}\Hat{\mathcal{D}}^2
   +4\Hat{\mathcal{F}}^2_{\mu\nu}.
\label{eq:(A3)}
\end{align}

\section{Small flow time expansion of the topological density}
\label{sec:B}
In this appendix, we present the small flow time expansion of an operator
corresponding to the topological charge density in the one-loop order. First,
to the quadratic order in the quantum field, we have
\begin{align}
   &\left.\epsilon_{\mu\nu\rho\sigma}
   G_{\mu\nu}^a(t,x)G_{\rho\sigma}^a(t,x)\right|_{O(b^2)}
\notag\\
   &=2\epsilon_{\mu\nu\rho\sigma}
   \left\{
   2\left[\Hat{\mathcal{D}}_\mu b_\nu(t,x)\right]^a
   \left[\Hat{\mathcal{D}}_\rho b_\sigma(t,x)\right]^a
   -b_\mu(t,x)\Hat{\mathcal{F}}_{\nu\rho}(x)b_\sigma(t,x)
   \right\}.
\label{eq:(B1)}
\end{align}
The contraction by the propagator~\eqref{eq:(2.38)} then yields
\begin{align}
   &\left\langle
   \left.\epsilon_{\mu\nu\rho\sigma}
   G_{\mu\nu}^a(t,x)G_{\rho\sigma}^a(t,x)\right|_{O(b^2)}
   \right\rangle_{\text{1PI}}
\notag\\
   &=-2g_0^2\epsilon_{\mu\nu\rho\sigma}
   \left[
   2\Hat{\mathcal{D}}_\mu^{ab}
   \left(\mathrm{e}^{2t\Hat{\Delta}}\frac{1}{\Hat{\Delta}}
   \right)_{\nu\rho}^{bc}
   \Hat{\mathcal{D}}_\sigma^{ca}
   +\Hat{\mathcal{F}}_{\mu\nu}^{ab}(x)
   \left(\mathrm{e}^{2t\Hat{\Delta}}\frac{1}{\Hat{\Delta}}
   \right)_{\rho\sigma}^{ba}\right]
   \left.\delta(x-y)\right|_{y=x}.
\label{eq:(B2)}
\end{align}
The same procedure as led to~Eq.~\eqref{eq:(3.16)} in the main text then
gives rise to
\begin{align}
   &\left\langle
   \left.\epsilon_{\mu\nu\rho\sigma}
   G_{\mu\nu}^a(t,x)G_{\rho\sigma}^a(t,x)\right|_{O(b^2)}
   -\left.\epsilon_{\mu\nu\rho\sigma}
   F_{\mu\nu}^a(x)F_{\rho\sigma}^a(x)\right|_{O(a^2)}
   \right\rangle_{\text{1PI}},
\notag\\
   &=-4g_0^2\epsilon_{\mu\nu\rho\sigma}
   \int_0^t\mathrm{d}\xi\,\xi^{-D/2}
   \int\frac{\mathrm{d}^Dp}{(2\pi)^D}\,\mathrm{e}^{-2p^2}
\notag\\
   &\qquad{}\times
   \tr\left[
   2\xi^{-1}
   \left(ip+\sqrt{\xi}\Hat{\mathcal{D}}\right)_\mu
   \left(\mathrm{e}^{4i\sqrt{\xi}p\cdot\Hat{\mathcal{D}}+2\xi\Hat{\Delta}}
   \right)_{\nu\rho}
   \left(ip+\sqrt{\xi}\Hat{\mathcal{D}}\right)_\sigma
   +\Hat{\mathcal{F}}_{\mu\nu}(x)
   \left(\mathrm{e}^{4i\sqrt{\xi}p\cdot\Hat{\mathcal{D}}+2\xi\Hat{\Delta}}
   \right)_{\rho\sigma}
   \right].
\label{eq:(B3)}
\end{align}
The expansion with respect to~$\xi$ is much simpler than~Eq.~\eqref{eq:(3.16)}.
Thus we give some details of the calculation for illustration.

First, in the integrand of~Eq.~\eqref{eq:(B3)}, any term that is symmetric
under the exchange of indices~$\{\mu,\nu,\rho,\sigma\}$ does not contribute
because of~$\epsilon_{\mu\nu\rho\sigma}$. Then, using~Eq.~\eqref{eq:(A1)}, it is
easy to see that the expansion of~Eq.~\eqref{eq:(B3)} to~$O(\xi^{-D/2+1})$ in
the integrand yields
\begin{align}
   &16g_0^2\epsilon_{\mu\nu\rho\sigma}
   \int_0^t\mathrm{d}\xi\,\xi^{-D/2+1}
   \int\frac{\mathrm{d}^Dp}{(2\pi)^D}\,\mathrm{e}^{-2p^2}
\notag\\
   &\qquad{}\times
   \tr\Bigl\{
   4p_\mu\left[
   \Hat{\mathcal{F}}_{\nu\rho}(x)p\cdot\Hat{\mathcal{D}}
   +p\cdot\Hat{\mathcal{D}}\Hat{\mathcal{F}}_{\nu\rho}(x)
   \right]\Hat{\mathcal{D}}_\sigma
   +4\Hat{\mathcal{D}}_\mu\left[
   \Hat{\mathcal{F}}_{\nu\rho}(x)p\cdot\Hat{\mathcal{D}}
   +p\cdot\Hat{\mathcal{D}}\Hat{\mathcal{F}}_{\nu\rho}(x)
   \right]p_\sigma
\notag\\
   &\qquad\qquad\qquad{}
   -2\Hat{\mathcal{D}}_\mu\Hat{\mathcal{F}}_{\nu\rho}(x)
   \Hat{\mathcal{D}}_\sigma
   -\Hat{\mathcal{F}}_{\mu\nu}(x)\Hat{\mathcal{F}}_{\rho\sigma}(x)\Bigr\}.
\label{eq:(B4)}
\end{align}
After the momentum integrations,
\begin{equation}
   \int\frac{\mathrm{d}^Dp}{(2\pi)^D}\,\mathrm{e}^{-2p^2}
   \begin{Bmatrix}
   1\\
   p_\mu p_\nu\\
   \end{Bmatrix}
   =\frac{1}{(8\pi)^{D/2}}
   \begin{Bmatrix}
   1\\
   \frac{1}{4}\delta_{\mu\nu}\\
   \end{Bmatrix},
\label{eq:(B5)}
\end{equation}
Eq.~\eqref{eq:(B4)} becomes
\begin{equation}
   \frac{16}{(8\pi)^{D/2}}
   g_0^2\epsilon_{\mu\nu\rho\sigma}
   \int_0^t\mathrm{d}\xi\,\xi^{-D/2+1}
   \tr\Bigl[
   \Hat{\mathcal{F}}_{\nu\rho}(x)\Hat{\mathcal{D}}_\mu\Hat{\mathcal{D}}_\sigma
   +\Hat{\mathcal{D}}_\mu\Hat{\mathcal{D}}_\sigma\Hat{\mathcal{F}}_{\nu\rho}(x)
   -\Hat{\mathcal{F}}_{\mu\nu}(x)\Hat{\mathcal{F}}_{\rho\sigma}(x)\Bigr].
\label{eq:(B6)}
\end{equation}
Finally, using Eq.~\eqref{eq:(3.17)}, we see that this combination identically
vanishes. We infer that, therefore, in the pure Yang--Mills theory,
\begin{equation}
   \epsilon_{\mu\nu\rho\sigma}G_{\mu\nu}^a(t,x)G_{\rho\sigma}^a(t,x)
   \stackrel{t\to0}{\sim}
   \left(1+0\cdot g_0^2\right)
   \epsilon_{\mu\nu\rho\sigma}F_{\mu\nu}^a(x)F_{\rho\sigma}^a(x)
   +O(t),
\label{eq:(B7)}
\end{equation}
to the one-loop order.

It turns out that, from very general grounds,
\begin{equation}
   \epsilon_{\mu\nu\rho\sigma}G_{\mu\nu}^a(t,x)G_{\rho\sigma}^a(t,x)
   \stackrel{t\to0}{\sim}
   \epsilon_{\mu\nu\rho\sigma}F_{\mu\nu}^a(x)F_{\rho\sigma}^a(x)+O(t)
\label{eq:(B8)}
\end{equation}
holds in the pure Yang--Mills theory in \emph{all orders of perturbation
theory\/}~\cite{Ce:2015qha}. To see this, one first notes~\cite{Luscher:2011}
\begin{equation}
   \partial_t\left[\epsilon_{\mu\nu\rho\sigma}G_{\mu\nu}^a(t,x)G_{\rho\sigma}^a(t,x)
   \right]=\partial_\mu W_\mu(t,x),\qquad
   W_\mu(t,x)
   =4\epsilon_{\mu\nu\rho\sigma}D_\lambda G_{\lambda\nu}^a(t,x)G_{\rho\sigma}^a(t,x),
\label{eq:(B9)}
\end{equation}
where $W_\mu(t,x)$ is a gauge-invariant dimension~$5$ axial-vector operator.
This shows that
\begin{equation}
   \epsilon_{\mu\nu\rho\sigma}G_{\mu\nu}^a(t,x)G_{\rho\sigma}^a(t,x)
   =\epsilon_{\mu\nu\rho\sigma}F_{\mu\nu}^a(x)F_{\rho\sigma}^a(x)
   +\partial_\mu\int_0^tdt'\,W_\mu(t',x).
\label{eq:(B10)}
\end{equation}
We then consider the small flow time expansion of~$W_\mu(t',x)$ in the last
term. Since there is no gauge-invariant axial vector of dimension~$<5$ in
the pure Yang--Mills theory, the small flow-time expansion of~$W_\mu(t',x)$
starts from a dimension~$5$ operator with an $O(t^{\prime0})$ coefficient
(possibly with logarithmic corrections). This implies that the last term
of~Eq.~\eqref{eq:(B10)} is~$O(t)$ and thus Eq.~\eqref{eq:(B8)}. Our explicit
one-loop calculation~\eqref{eq:(B7)} is consistent with this general
property~\eqref{eq:(B8)}, as it should be.

\end{document}